# Josephson vortex system in a flux-flow regime in electron doped high-$T_c$ superconductor $Nd_{2-x}Ce_xCuO_4$

T. B. Charikova, A. M. Bartashevich, V. N. Neverov, M. R. Popov, N. G. Shelushinina and A.A Ivanov

**Abstract**

For an epitaxial film of an electron-doped superconductor $Nd_{2-x}Ce_xCuO_4$ ($x = 0.145$), in a mixed state, a well-defined step structure is observed on the dependence of the $c$-axis flux-flow resistance on the magnetic field parallel to $CuO_2$ layers. It is shown that in the region of a free flow of Josephson vortices, the structure is periodic with a period of $\Delta B \sim \Phi_0/(\lambda_{ab}\lambda_c)$. It is essential that the magnetic penetration depths in the ab-plane ($\lambda_{ab}$) and along the $c$-axis ($\lambda_c$) determine the sizes of the Josephson vortices in the Lawrence-Doniach model for an anisotropic layered superconductor.

*Keywords*: Intrinsic Josephson junction; Josephson vortex; flux-flow resistance, magnetic penetration depth, Lawrence-Doniach model

## 1. Introduction

Oxide superconductors are layered compounds with building blocks consisting of conductive $CuO_2$ layers separated by buffer layers that serve as charge reservoirs (see monographs [1], [2], [3], [4] for a detailed description). Highly anisotropic high-temperature superconductors (HTSCs) can be considered as a "package" of superconducting $CuO_2$ layers coupled by Josephson interactions [5], [6], [7], [8]. The new properties of these materials compared to single Josephson junctions are associated with their multilayer structure and the atomic thickness of the superconducting layers.

Because of the layered structure and the resulting strong anisotropy, a high-temperature cuprate crystals in the superconducting state can be considered as a stack of tunneling Josephson junctions. These junctions are called intrinsic Josephson junctions (IJJ) [9], [10]. The intrinsic Josephson effect as tunneling of Cooper pairs between adjacent $CuO_2$ planes inside a highly anisotropic layered HTSC is an integral part of many theories on this subject (see, e.g., reviews [9], [10] and numerous references therein).

Research in a magnetic field constitutes a vast and rapidly developing field of study of layered HTSCs, promising for practical applications. The key phenomenon associated with the intrinsic Josephson effect is the oscillations of various properties of superconducting tunnel junctions with a magnetic field.

Let us highlight measurement series of dc-resistivity in the presence of the c-axis current with magnetic field parallel to the ab-plane (the flow resistance of Josephson vortices (JVs)) for a micron sized mesoscopic structures made of single crystalline $Bi_2Sr_2CaCu_2O_{8+\delta}$ (Bi2212) [11], [12], [13], [14] and also for annealed $YBa_2Cu_3O_x$ intrinsic Josephson junctions having high anisotropy [15] and for pnictide $(Sr_4V_2O_6)Fe_2As_2$ with intrinsic Josephson effect [16].

The novel oscillations of the flux-flow resistance as a function of magnetic field have been found. The period of the oscillations depends only on the interlayer distance, $s$, and the width of the junction, $L$, perpendicular to the external field. The observed period of oscillations, agrees well with the increase of one fluxon per two junctions ($Hp=\Phi_0/2Ls$), may correspond to formation of a triangular JV lattice. Then these oscillations may originate from commensuration between the triangular vortex lattice and the width of the junction perpendicular to the field.

It is observed that this oscillating period becomes doubled above a certain field, indicating the structure transition from triangle to square structure. It was interpreted as the effect of the edge deformation of the JV lattice due to surface current of intrinsic Josephson junctions as pointed by Koshelev [17], [18]. According to Koshelev this unusual twice oscillation period can be interpreted as the extension of the deformation area of the Josephson vortex arrangement over the edges, which is essential for the phenomenon.

By using numerical simulation technique, the similar results but claiming more explicitly the formation of the triangular lattice in the twice $\Phi_0$ oscillation period region and the square lattice with the $\Phi_0$ oscillation region have been obtained by Machida [19].

We report the observation of a periodic structure (quantized steps) in the curve of the dependence of $c$-axis resistance, $R_c$, on magnetic field, $B$, parallel to the $CuO_2$ layers, in a mixed state of the epitaxial film $Nd_{2-x}Ce_xCuO_4/SrTiO_3$ with $x = 0.145$. In Section 2 we provide a brief summary of the basic Josephson effect formulas both for a single junction and for a multilayer system in order to place our experiments in the series of results on Josephson junctions.

## 2. Theoretical considerations

The Josephson effect is one of the most striking and practically important phenomena in the physics of superconductors. By applying the microscopic theory to a tunnel junction between the superconductors Josephson in 1962 [20] made two unforeseen predictions of effects that became known as the dc and ac Josephson effects. Josephson was the first to predict the tunneling of superconducting Cooper pairs. For this work he was awarded the Nobel Prize, 1973.

The Josephson effects, from the first theoretical work to contemporary studies in the foundations of superconducting computers and supersensitive magnetometers are described in numerous monographs and textbooks (see, for example, [3], [21], [22], [23]). The experiments on quantum interference serve as a clear evidence of the fact that the value of the wave function phase of superconducting electrons is observable. It is shown that superconductivity is a macroscopic quantum phenomenon.

*2.1. Formulas for a single Josephson junction*

- *dc Josephson effect*. The static (dc) effect corresponds to a supercurrent flow between the ground state of two superconductors in an electron tunneling experiment with the absence of an externally applied potential difference. The first observations of the dc Josephson effect were made by Anderson and Rowell [24] and by Rowell [25] in 1963.

Josephson [20] demonstrated that a non-damping superconducting current, $I_s$, which can flow through a tunnel junction between two superconductors (a contact between superconductors through a thin layer of insulator, SIS contact), in the absence of an externally applied potential difference, is described by the expression

$$I_s = I_c \sin\varphi. \tag{2.1}$$

Here $I_c$ is the critical current (the maximum value of supercurrent that a Josephson junction can support), φ is the phase difference of the order parameter (of the Ginzburg-Landau wave function) in the two superconductors which form the banks of the junction.

- *ac Josephson effect*. The dynamical (ac) Josephson effect considers alternating current flow between the ground states of two superconductors when there is a small potential difference $V$ between them. The current flow is accompanied by an emission of electromagnetic waves, the frequency $v$ of which is given by $hv = 2eV$. This phenomenon, called *Josephson generation*, has found indirect experimental support by Shapiro [26] and Esk et al. [27] and then the existence of the ac effect was confirmed in a straightforward refined experiment by Ivar Giaever [28] (Nobel Prize, 1973) and in experiments of Yanson et al. [29].

The fundamental Josephson equations with a fixed voltage $V$ across the junction have the appearance [21]:

$$I(t) = I_c \sin \varphi(t) \qquad (2.3)$$

$$\frac{\partial \varphi}{\partial t} = \frac{2eV}{\hbar} \qquad (2.4)$$

Thus, the phase will vary linearly with time and the current through the Josephson junction will be a sinusoidal ac (alternating current) with amplitude $I_c$ and frequency $\omega_J = 2eV/\hbar$.

- *Ferrel-Prange equation.* We'll look into the response of the SIS system to an external magnetic field parallel to the plane of the junction. Let the $x$-axis lie in the plane of the junction, and the magnetic field is directed along the $y$-axis. The region along the $z$-axis, where the current flows and the magnetic field exists, has the size $d$ (see Fig. 2.1). Here $d = 2\lambda_L + s$ with $\lambda_L$ being the London penetration depth, $s$ being the thickness of the insulating layer of the tunnel junction.

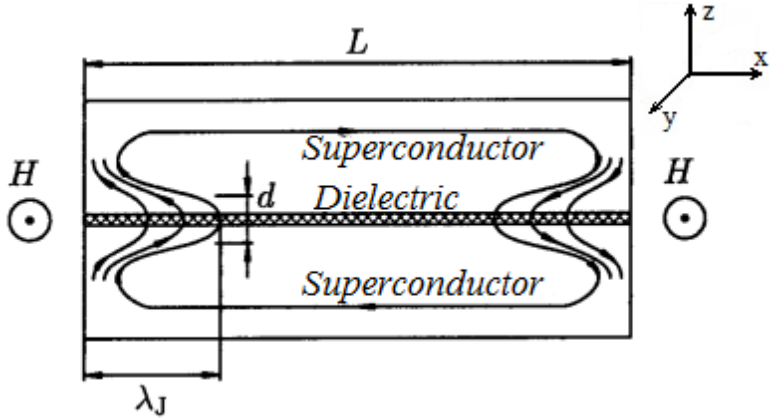

Fig. 2.1. Josephson tunnel junction placed in a magnetic field $H$. The distribution of the screening Meissner current is shown (after [22]).

In the presence of a magnetic field in the described geometry, the phase difference between the two superconductors, $\varphi$, becomes dependent on the $x$ coordinate. The dependence $\varphi(x)$ is mathematically described by the equation [30]:

$$\frac{d^2\varphi}{dx^2} = \frac{1}{\lambda_J^2} \sin \varphi, \qquad (2.5)$$

where $\lambda_J$ (in SI system) is equal to

$$\lambda_J = \left(\frac{\Phi_0}{2\pi\mu_0 j_c d}\right)^{1/2}. \tag{2.6}$$

Here vacuum permeability $\mu_0 = 4\pi \cdot 10^{-7} H/m$, $\Phi_0 = h/2e = 2.07 \cdot 10^{-15} Wb$ is the magnetic flux quantum and $j_c$ is the critical current density, expressed in $A/m^2$, $d$ is taken in meters. The quantity $\lambda_J$ represents the penetration depth of magnetic field into a Josephson junction (*Josephson penetration depth*). Note that the relation $\lambda_J \gg \lambda_L$ is valid, which is consistent with the definition of "weak superconductivity" (a term introduced by Anderson [31] for the Josephson junction).

- *"Small" Josephson junctions.* Let the size of the junction be $L \ll 2\lambda_J$, then it has an uniform distribution of magnetic field $H$ (see Fig. 2.1) and in the stationary case we have:

$$\frac{d\varphi}{dt} = 0; \quad \frac{d\varphi}{dx} = \frac{2\pi d}{\Phi_0} H.$$

Integration of $d\varphi/dx$ gives

$$\varphi(x) = \frac{2\pi d}{\Phi_0} H + \varphi_0 \tag{2.7}$$

and substituting (2.7) into (2.1), we find

$$I_s = I_c \sin\left(\frac{2\pi d}{\Phi_0} H + \varphi_0\right), \tag{2.8}$$

which indicates that the tunneling supercurrent is spatially modulated by the magnetic field.

After some conversions you can get that the maximum dissipation-free total current through the junction, $I_{max}$, is determined by the expression

$$I_{max} = I_c \left|\frac{\sin(\pi\Phi/\Phi_0)}{\pi\Phi/\Phi_0}\right| \tag{2.9}$$

with $\Phi$ being the total magnetic flux in the Josephson junction. In the simplest case $\Phi = HLd$ (see Fig. 2.1) and, in general, the expression for the total magnetic moment depends on the geometry of the sample (rectangular junction, annular junction, double junction, etc.)

Due to the periodic character of the expression (2.9), situations can be realized in which the net tunneling current is zero. In particular, a rectangular junction with a uniform zero field tunneling current distribution exhibits a dependence of the maximum supercurrent on the applied magnetic field in the form of a *Fraunhofer-like diffraction pattern* (see, for example, Fig. 6.9 in

[3] or Figs 24.5 and 24.6 in [22]). The first observation of this effect was made by Rowell in 1963 [25].

The extremely high sensitivity of the Josephson current to the magnetic field is the key point to many important applications of the Josephson effects. In particular, this served as the basis for creating unique magnetometers known as *SQUIDs* (Superconducting Quantum Interference Devices).

- *Long Josephson junctions (JJs).* A long JJ (Fig. 2.2) is a junction which has one dimension (say *x*) long with respect to the Josephson penetration depth, $\lambda_J$. In general, the phase difference is a function of the time and the spatial coordinate: $\varphi(x,t)$.

In the simplest approach, the electrodynamics of the junction is described by the following *sine-Gordon equation* [21], [22], [32]:

$$\lambda_J^2 \frac{\partial^2 \varphi}{\partial x^2} - \frac{1}{\omega_p^2} \frac{\partial^2 \varphi}{\partial t^2} = \sin \varphi \qquad (2.10)$$

with $\lambda_J$ being the *Josephson penetration depth* and $\omega_p = \sqrt{2\pi I_c / \Phi_0 C}$ being the *Josephson plasma frequency* of the system, *C* is the capacitance per unit of length of the junction.

The equation (2.10) is an unperturbed version of sine-Gordon equation which does not contain any dissipative or bias terms. This very important equation is widely used throughout applied science.

One of the solutions of the equation (2.10) has the form of a solitary excitation, with the peculiar character of a particle-like or "soliton" solutions:

$$\varphi(x) = 4 \arctg \exp(x/\lambda_J). \qquad (2.11)$$

The soliton solution of the sine-Gordon equation has a well-defined physical meaning, it corresponds to a quantum of flux trapped in the junction and is usually referred to as *fluxon* or *Josephson vortex* in the context of long JJ.

Figure 2.2 schematically shows the Josephson vortex in the junction and the form of $\varphi(x), \frac{d\varphi}{dx} \sim H$ and $\frac{d^2\varphi}{dx^2} \sim j_s$. It is visible that the phase changing from 0 to $2\pi$ along the junction takes place.

The width scale for the change of the phase, the width of the vortex, is given by the Josephson penetration depth $\lambda_J$. A spatial variation of the phase gives account of a magnetic flux in the system, whose total integral is equal to one flux quantum $\Phi_0$. In addition, it follows from the Josephson relations that a nonzero phase variation corresponds to a flowing supercurrent which, as shown in the figure, circulates around the vortex core.

It turns out that the behavior of a Josephson junction in an external magnetic field is in many ways reminiscent of the behavior of a type II superconductor: when the external field exceeds the characteristic (lower critical) field, superconducting vortices carrying a quantum of magnetic flux $\Phi_0$ will begin to penetrate into the junction. But, in contrast to the Abrikosov vortex in type II superconductor, the Josephson vortex does not have a normal core as the amplitude of the order parameter $|\Psi(x)|$ (and, consequently, the density of superconducting carriers, $n_s = |\Psi(x)|^2$) remains constant at the center of the Josephson vortex.

Similarly to single fluxons, the Eq. (2.10) also supports for multifluxon solutions formed by a set of fluxons or antifluxons, or combinations of both. In addition to static solutions ($\varphi = 0$ and static fluxons), the unperturbed sine-Gordon equation also supports a family of dynamical solutions including moving fluxons [32].

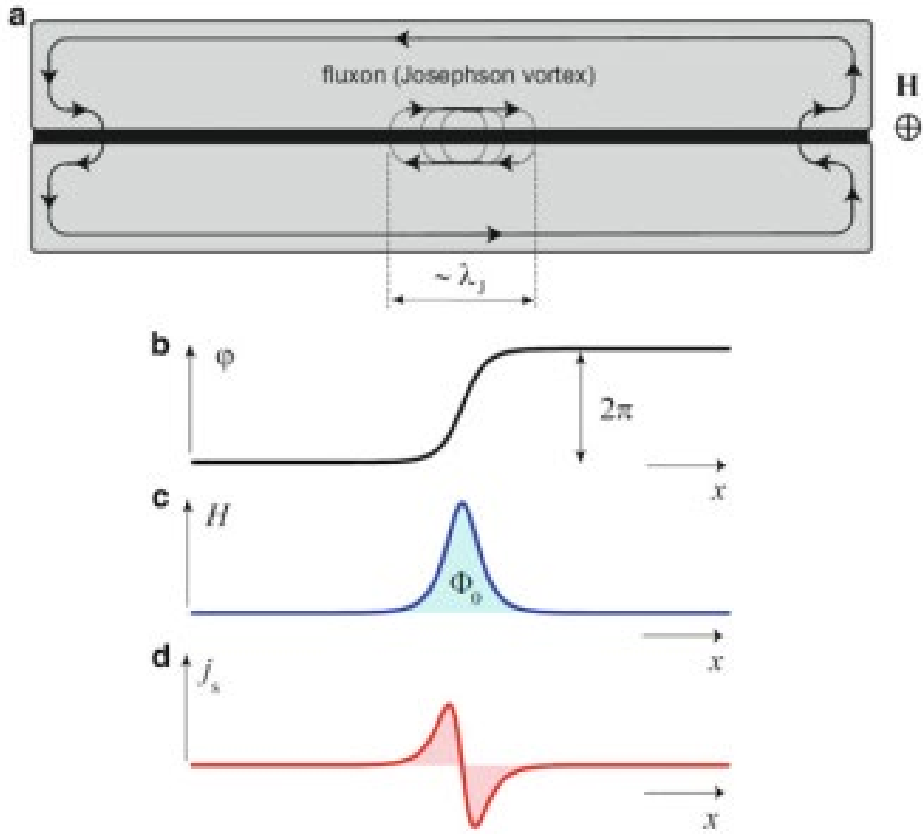

**Fig. 2.2. (a)** Schematic diagram of a long JJ with a vortex in it, **(b)** phase $\varphi$ **(c)** field $H \sim d\varphi/dx$ and **(d)** density of superconducting current $j_s \sim \sin \varphi$ for a soliton in the junction. The vortex extends over a physical distance of the order of $\lambda_J$ (after [32]).

*2.2. Josephson vortices in layered superconductors*

In a layered superconductor, the structure of the individual vortices as well as that of the vortex lattice strongly modified (see review of Blatter et al. [8], works of Clem and Coffey [33] and of Koshelev and Dodgson [34] and references therein).

As already said, layered superconductors are materials made from a stack of alternating thin superconducting layers separated by non-superconducting regions of width $s$. If the interlayer distance $s \gg \zeta_c$ ($\zeta_c$ being the correlation length of the Cooper pair along the c axis) the superconducting layers are essentially two-dimensional (2D) as there is no variation in fields, or in the superconducting order parameter, across each layer. Such a superconductor can carry super-currents along the layers, as well as between the layers. This is due to the Josephson tunneling of Cooper pairs [20] across the insulating regions that separate neighboring superconducting layers, i.e., each pair of neighboring layers forms one Josephson junction. In general, the *c*-axis (Josephson) supercurrents are weaker than the supercurrents along the layers.

- *Structure of Josephson vortex.* Consider the properties of Josephson vortices generated by the magnetic field applied in the layer's direction. The theoretical description is based on the Lawrence-Doniach model [5]. In that case the Josephson interlayer currents are involved in the formation of vortices.

The structure of the supercurrent distribution around the Josephson vortex in an anisotropic high-*Tc* superconductor was calculated by Clem and Coffey [33]. The characteristic sizes of a Josephson vortex in such a superconductor have two length scales, $\lambda_{ab}$ (penetration depth of the magnetic field in the in-plane direction) and $\lambda_c$ (magnetic penetration depth in the *c*-axis direction), both much larger than the thickness of the intrinsic Josephson junction. The streamlines of the supercurrent around the vortex center, which also represent contours of constant magnetic field, are elliptical except for the zigzags due to the intervening insulating layers (see Fig. 2.3).

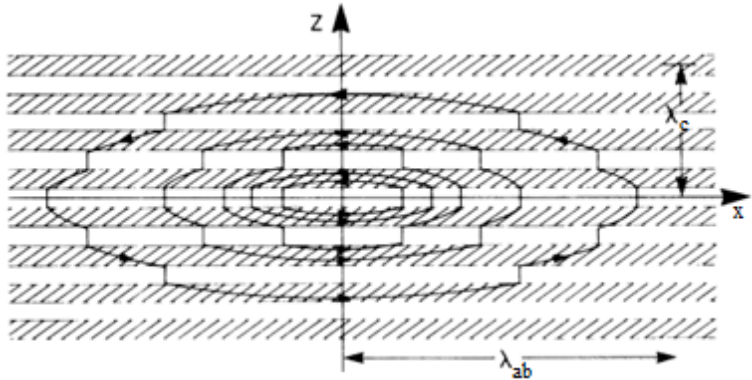

Fig. 2.3. Sketch of the supercurrent distribution around a single vortex in the barrier region of the central Josephson junction in an infinite layer model of an anisotropic high-$T_c$ superconductor. The vortex is parallel to the *ab* planes (*y* direction of magnetic field). The magnetic penetration depths $\lambda_{ab}$ and $\lambda_c$ give the scale for the decay of the supercurrent in the *x* and *z* (c- axis) directions, respectively (after [33]).

- *Core structure.* Josephson vortices do not have normal cores, in contrast with Abrikosov vortices in standard anisotropic superconductors. Rather, the centers of these vortices lie between layers, where the Josephson character of the interlayer current is important, and the normal core is replaced by the nonlinear core, the region within which the phase difference between the two central layers sweeps from 0 to $2\pi$ [8], [34].

The core size is given by the lengths $s$ along the $c$ axis and $\gamma s$ along the $ab$ plane (i.e., along layers), respectively. Here $\gamma$ is the anisotropy parameter, $\gamma = \lambda_{ab}/\lambda_c = \sqrt{\rho_c/\rho_{ab}}$.

At low fields the Josephson vortices are isolated and form a triangular lattice, strongly stretched along the direction of layers. When magnetic field exceeds the crossover field, $B_{cr} = \Phi_0/\pi\gamma s^2$, the cores of Josephson vortices start to overlap and a dense Josephson lattice begins to form ([34] and references therein). In fields $B \gg B_{cr}$ all interlayer spacings are filled by vortices and the vortex lattice in a given interlayer spacing becomes very similar to that in a single Josephson junction.

It's necessary to note that the Josephson vortex in a layered superconductor is characterized by two length scales, describing the core ($s$, $\gamma s$) and the magnetic size ($\lambda_c$, $\lambda_{ab}$) of the Josephson vortex. This is in contrast with the case of a Josephson vortex in a single junction between two weakly coupled bulk superconductors, for which there is only one characteristic length scale along the $x$ axis, namely, the Josephson penetration depth $\lambda_J$ (see Eq. (2.6) and Fig. 2.2).

## 3. Sample characteristics and equipment

The magnetotransport properties of $Nd_{2-x}Ce_xCuO_4$ system with $x = 0.145$ were investigated. The sample is the $Nd_{2-x}Ce_xCuO_4/SrTiO_3$ epitaxial film grown in such a way that the $c$-axis of the $Nd_{2-x}Ce_xCuO_4$ lattice (orientation ($1\bar{1}0$)) is directed along the long side of the $SrTiO_3$ substrate [35] (Fig. 3.1).

The measurements were carried out on the Quantum Design PPMS 9 and in the original certified setup for measuring of galvanomagnetic effects with the solenoid "Oxford Instruments" (Center for Nanotechnologies and Advanced Materials, IFM UrB RAS) in magnetic fields up to 9 T at helium temperatures, $T = 1.9$ K and 4.2 K. The external electric current, $J$, is applied along the $c$-axis, across the $CuO_2$ planes ($z$ direction), the magnetic field vector, $B$, is perpendicular to the substrate plane, lies in the $CuO_2$ plane with $B \perp J$ ($y$ direction). In the flux – flow regime (see below) the vortices move perpendicular to the current and to the magnetic field in the $x$ direction.

The coordinate system used here (in accordance with Section 2) is shown in Fig. 3.1 at the top right.

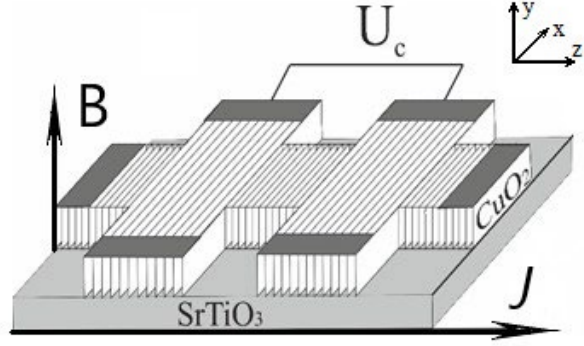

Fig. 3.1. Sample $Nd_{2-x}Ce_xCuO_4/SrTiO_3$, $x = 0.145$, c-axis is parallel to the long side of the film.

In our previous works [36], [37], we have found that in the concentration range $x = 0.145$ and $0.15$ $Nd_{2-x}Ce_xCuO_4$ compound is in state of a 2D metal with metallic in nature conductivity in the $CuO_2$ layers and non-metallic conductivity across the layers. The resistance across the $CuO_2$ layers, $\rho_c$, is significantly higher than the resistance, $\rho_{ab}$, in the conductive $CuO_2$ layers and $\rho_c(T)$ reveals a non-metallic temperature dependence for all films studied. It indicates the quasi-two-dimensional character of conductivity in our samples.

As it is seen from (Fig.3.2 a) the resistivity anisotropy coefficient, $\rho_c/\rho_{ab}$, reaches the values of $10^2 - 10^3$. For composition $x = 0.145$ we have $T_c = 15.7$ K and $\rho_c/\rho_{ab} = 290$.

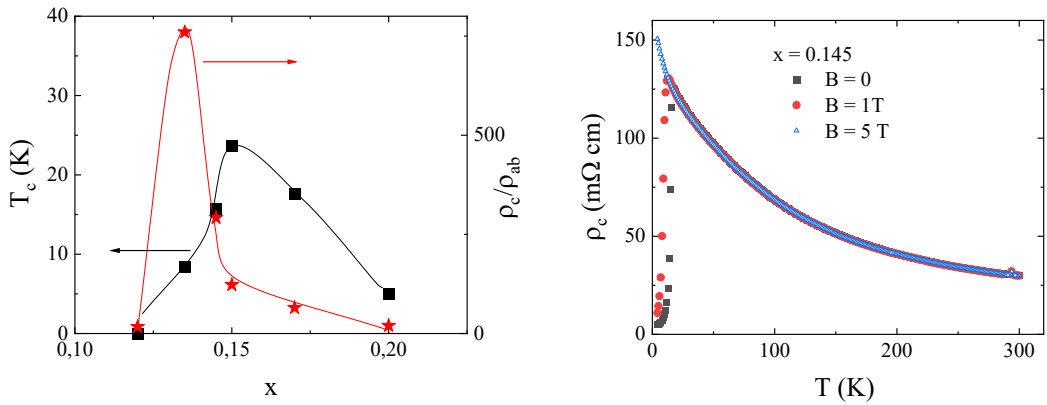

Fig. 3.2. (a) Dependences of the critical temperature, $T_c$, and the resistivity anisotropy coefficient, $\rho_c/\rho_{ab}$, on dopant Ce content ($x$) (b) $\rho_c(T)$ dependences in different magnetic fields. The magnetic field **B** is parallel to the $CuO_2$ planes (perpendicular to the current **J**).

In the sample we studied in the normal state we clearly observe a sharp (almost exponential) increase in $\rho_c(T)$ with a decrease in $T$, as $T \to T_c$ (Fig. 3.2 b). It is generally accepted that the observed behavior of $\rho_c(T)$ is a manifestation of a tunnel conduction between $CuO_2$ layers through the insulating buffer barriers. The magnetic field **B,** directed perpendicular to the plane of the sample, leads to a gradual decrease of $T_c$ and to a suppression of superconductivity at $B = 5$ T (see Fig. 3.2 b).

In conclusion of this section, we present the values of the system parameters necessary for the analysis of experimental data in a flux-flow regime (see Table 1). Here $L$ is the width of the sample (dimension in the $x$ direction), $t$ is the thickness of the film (dimension in the $y$ direction), $s$ is the distance between the $CuO_2$ layers, $\xi_c$ is the correlation radius of the Cooper pair along the $c$ axis ($z$ direction), $\lambda_{ab}$ and $\lambda_c$ are the penetration depth of the magnetic field along the $ab$ planes and along the $c$ axis, respectively. To assess the magnetic penetration depth $\lambda_{ab}$ in $Nd_{2-x}Ce_xCuO_4$ we used the results of measurement by Prozorov [38] for $Pr_{2-x}Ce_xCuO_4$ ($\lambda(0) = (2790 \pm 150)$Å).

Further, $\gamma = \lambda_{ab}/\lambda_c = \sqrt{\rho_c/\rho_{ab}}$ is the anisotropy parameter, $d = s + 2\lambda_c$ is the Josephson vortex size in the $z$ direction and $\lambda_J$ is the Josephson penetration depth, the $x$ direction vortex size for a single Josephson junction (see Section 2, Eq. (2.6)).

Table 1. Parameters of investigated $Nd_{2-x}Ce_xCuO_4$ film.

| x | $L, \mu m$ | $t, nm$ | $s, nm$ | $\xi_c, nm$ | $\lambda_{ab}, nm$ | $\lambda_c, nm$ | $\gamma$ | $d, nm$ | $\lambda_J, \mu m$ |
|---|---|---|---|---|---|---|---|---|---|
| 0.145 | 800 | 400 | 0.6 | 0.23 | ~300 | ~16 | 18 | ~33 | 100 |

Looking at the table, we note the following important relationships of the parameters:

- For $\xi_c < s$ the discreteness of the structure becomes relevant, and an appropriate description of it is in terms of a set of weakly coupled superconducting (quasi-2D) layers, which is provided by the model of Lawrence and Doniach [5].

- With $\lambda_J \ll L$ we would be dealing with a *long Josephson junction* (see Sec. 2.1) if the structure under study could be considered as a set of single junctions (for $d \cong s + 2\xi_c$).

- But we have another ratio of the parameters: $d \gg s + 2\xi_c$ ($d/(s + 2\xi_c) = 20 \div 30$) and the behavior of the system in a magnetic field parallel to the $CuO_2$ planes (in particular, the structure of Josephson vortices) must be considered within the framework of a multilayer (Josephson-coupled layer) model which is based on the Lawrence-Doniach model in the London limit [8], [33], [34] (see Sec. 2.2).

## 4. Experimental results and discussion

### *4.1 Current-voltage characteristics at B=0*

After effective optimal annealing, the $CuO_2$ layers in a cuprate crystal are well separated from each other by buffer layers (as evidenced by the sharp increase of $\rho_c(T)$ in the normal state at $T \to T_c$). Therefore, all conditions exist for the formation of a system of intrinsic Josephson junctions (IJJs) of the SIS-type after the sample transition to the SC state.

The intrinsic Josephson effect, conditioned by the tunneling of charge carriers in both the superconducting and the normal states of cuprate multilayer HTSCs, has been intensively studied in recent decades. The distinctive structure of the current-voltage characteristics for current in the *c*- direction, with a large number of hysteresis resistive branches, is most pronounced in the highly anisotropic Bi- and Tl-systems [39], [40] (see also reviews [9], [10] and references therein).

Really, the I-V- characteristic demonstrates multiple branches in the resistive state of our sample (Fig. 4.1). This behavior corresponds to the standard superconductor-insulator–superconductor (SIS) tunnel junction and characterizes the $Nd_{2-x}Ce_xCuO_4$ compound as a system with intrinsic Josephson junctions (for a detailed discussion, see our previous work [41]).

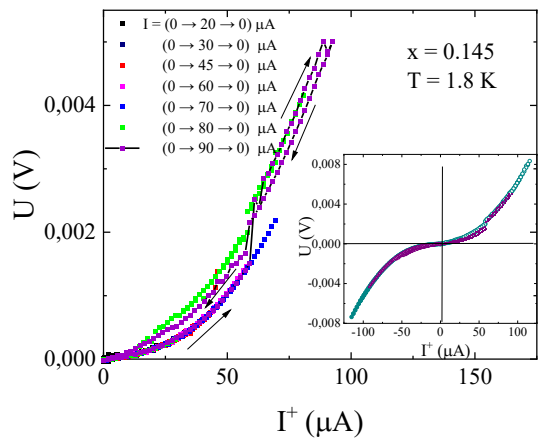

Fig.4.1. Hysteresis I-V dependence and jumps from one resistive branch to other for $Nd_{2-x}Ce_xCuO_4/SrTiO_3$ epitaxial film ($x = 0.145$) at $T = 1.8$ K. The common I-V- characteristic of this type of films shown on the insert.

Each junction has a hysteresis I-V- characteristic that one can see on the Fig.4.1 for the optimally annealed $Nd_{2-x}Ce_xCuO_4$ /$SrTiO_3$ epitaxial films with $x = 0.145$ at $T = 1.8$ K. Our precise I-V- measurements made it possible to visualize the jump from one resistive branch to other with an increase in current and back from one to other with a decrease in it.

Our experimental measurements of the critical current temperature dependence on the optimally annealed $Nd_{2-x}Ce_xCuO_4$ /$SrTiO_3$ epitaxial films ($x = 0.145$) (Fig. 4.2) confirm linear dependence $I_c(T)$, which also proves the existence of the SIS-type Josephson tunnel junctions in the compound $Nd_{2-x}Ce_xCuO_4$.

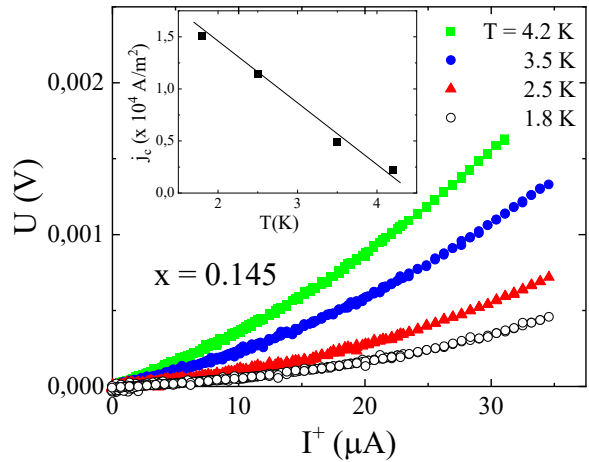

Fig. 4.2. The first I-V branch for optimally annealed $Nd_{2-x}Ce_xCuO_4$ /$SrTiO_3$ epitaxial film ($x = 0.145$) at different $T$ with temperature dependence of the critical current density, $j_c(T)$, on the insert.

### *4.2. Quantum stair-step flux – flow magnetoresistance*

The magnetotransport properties in a flux – flow regime for $Nd_{2-x}Ce_xCuO_4$/$SrTiO_3$ ($x = 0.145$) epitaxial films, with the c-axis directed along the long side of the film, in the geometry shown in Fig. 3.1, were researched. Measurements of $R_c(B)$ (resistance between $CuO_2$ planes) were carried out on the PPMS setup in the range of magnetic fields $B = (-9 - +9)$ T at temperatures $T = 1.9$ and $4.2$ K.

The PPMS installation allows for measurements with great accuracy, in high magnetic fields and at the current $I = 100$ µA. The common view of the dependence of resistance $R_c$ on magnetic field $B$ for investigated sample is presented on Fig.4.3 for $T = 1.9$K.

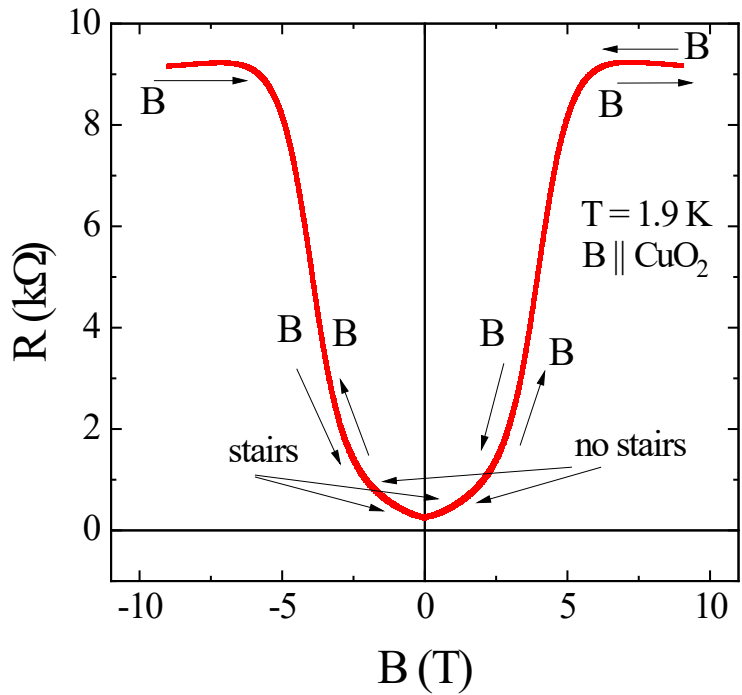

Fig.4.3. The dependence $R_c(B)$ for the sample studied.

Let us note that in fields $B = (-2.75 - +2.75)$ T resistance steps are observed. We start measuring the resistance at $T = 1.9$ K in a magnetic field of $B = 9$ T, decrease the magnetic field value, and uniform steps appear on the $R_c(B)$ dependence in a field of $B = 2.75$ T. After the magnetic field has decreased to zero, we begin to increase the field in the opposite direction ($B+$). In this case, the steps on the $Rc(B)$ dependence are not fixed. We increase the magnetic field value up to $B = 9$ T, after which we begin to decrease the magnetic field value again. Uniform steps appear on the $R_c(B)$ dependence in a magnetic field of $B = +2.75$ T (Fig. 4.4)

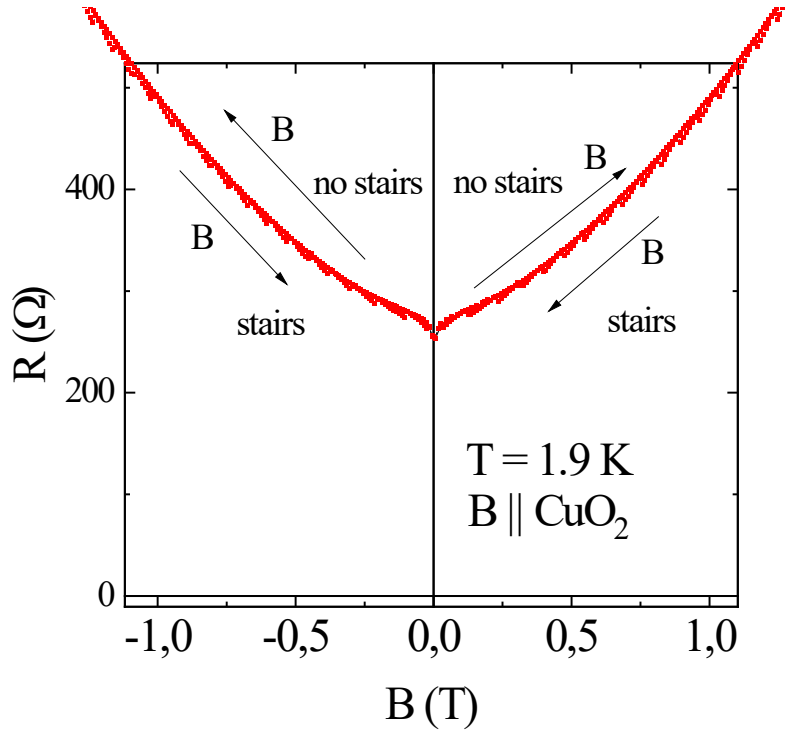

Fig. 4.4. $R_c(B)$ dependencies for the sample studied in the magnetic field range $-1T < B < +1T$ ($B \parallel CuO_2$ layers).

## *4.3. Analysis of experimental data*

As is known (see references in sections 2.1 and 2.2), at $B > B_{c1}$ (see definition of $B_{c1}$ below), in a mixed state, the magnetic field penetrates into the superconducting Josephson system (a single JJ or the stacks of JJs) in the form of magnetic flux vortices. In layered cuprates, if the magnetic field is applied along the $CuO_2$ layers, these are Josephson vortices.

Turning on the transport current ***J***, perpendicular to the layers, with ***J*** ⊥ ***B***, leads to the movement of vortices across the sample when they arise at one edge of the sample and escape at the opposite side. It is the flux-flow regime. Moving Josephson vortices generate both in-plane and inter-plane electric fields, which induce dissipative quasiparticle currents and the appearance of the flux-flow resistivity, $\rho_f$.

In the geometry presented in Fig. 3.1, when ***B*** $\parallel CuO_2$ layers, at $T < T_c$ and $B \leq 6T$, in our sample we are dealing just with the flux-flow regime for the Josephson vortices. The situation geometrically corresponds to theoretical calculations of monotonic dependence $\rho_f(B)$ by Clem and

Coffey [33], as well as by Koshelev [42]: the current $\mathbf{J}$ is directed along the $c$-axis, $\mathbf{B} \perp$ the $c$-axis, $\mathbf{B} \perp \mathbf{J}$.

In [33] only dissipation due to the tunneling of quasiparticles between the layers was taken into account in calculations of the viscosity coefficient and in [42] a case of dominating in-plane dissipation channel was examined. Overall, according to [33] and [42] a field dependence of the flux-flow resistivity $\rho_f(B)$ one can imagine in the form:

$$\frac{\rho_f}{\rho_c} = \alpha \frac{B}{(\Phi_0/s^2)} + \beta \frac{B^2}{B^2 + B_\sigma^2} \qquad (4.1)$$

with $\rho_c$ being the normal state resistivity along the $c$-axis.

Thus, the indicated dependence $\rho_f(B)$ should have the linear field dependence in a weak magnetic field, then, for strong in-plane dissipation, should have pronounced upward curvature and approaches to saturation on $\rho_c$ at $B > B_\sigma$, as shown schematically in Fig. 1 of Ref.42.

The monotonous behavior component of $R_c(B)$ observed for the studied sample, as shown in Fig. 4.3 for $T = 1.9$K, qualitatively corresponds to the dependence of the type (...) with $B_\sigma \approx 6T$.. However, the presence of periodic (quasi-periodic) structures such as oscillations or steps on the curve $R_c(B)$ is already a manifestation of further interaction of the current flowing in the Josephson structure with certain magnetic excitations of the system (for example, with vortices).

Let us analyze the observed stepwise dependence of $R_c(H)$ in our sample. Fig. 4.4 shows the experimental results for $R_c(B)$ at one direction of the magnetic field ($B^+$) for both the ascending branch (field input, $B_\uparrow$) and the descending branch (field output, $B_\downarrow$) and also the difference, $\Delta R_c$, in the values of $R_c(B_\downarrow)$ and $R_c(B_\uparrow)$. Subtracting the smooth $R_c(B_\uparrow)$ dependence from the step $R_c(B_\downarrow)$ one almost eliminates its monotonic part. As a result, we have an oscillatory-like pattern, shown in Fig. 4.4 b.

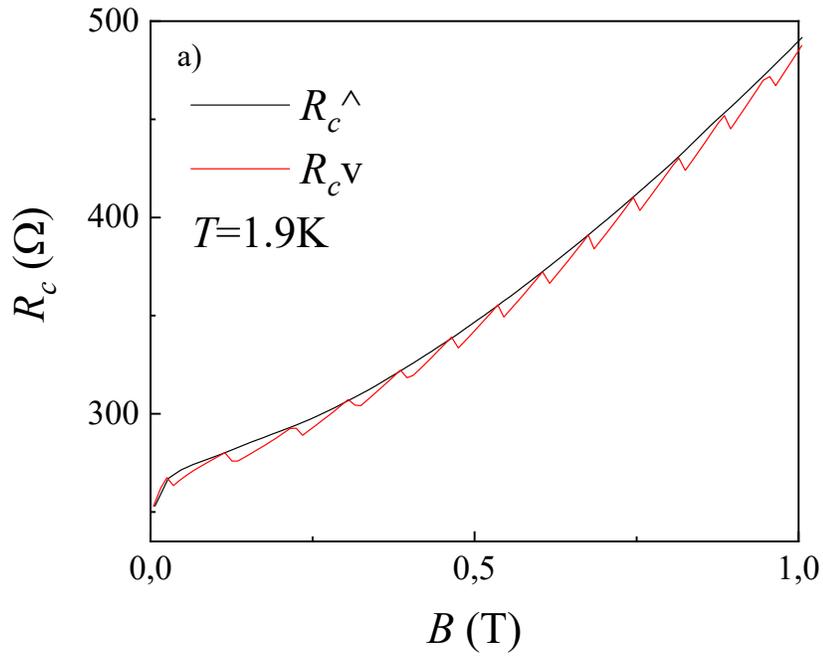

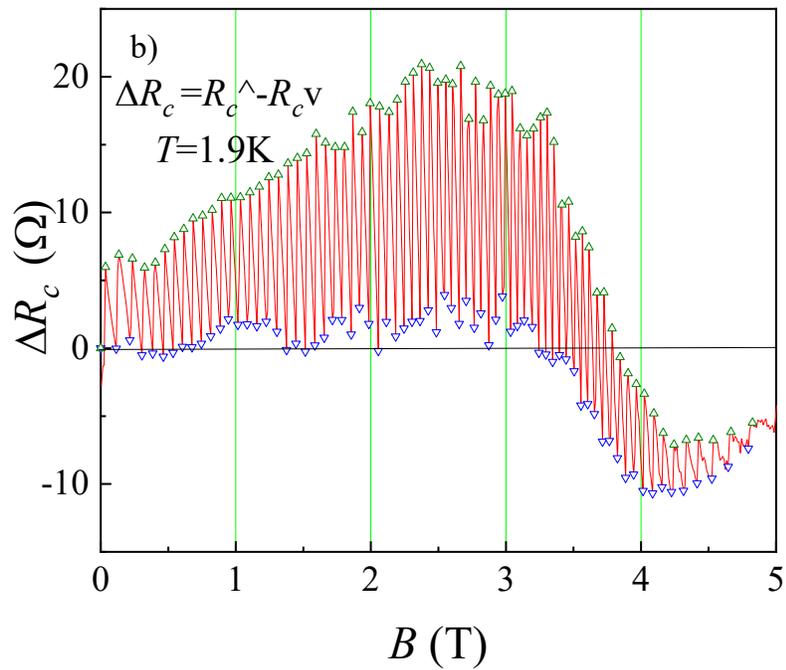

Fig. 4.4. (a) Dependencies of $R_c(B)$ for positive values of the magnetic field at the input ( $B_\uparrow$, blue line) and at the output ($B_\downarrow$, red line) of the field (fragment) and (b) the difference $\Delta R_c = \{R_c(B_\uparrow) - R_c(B_\downarrow)\}$, where the minimum points are marked in blue and the maximum points are marked in green.

The width of the steps in dependence on the peak number and on the corresponding magnetic field strength and the amplitude of bursts ("oscillations") in dependence on the magnitude of the magnetic field are presented in Figs 4.5 a,b. It can be seen that the amplitude of the peaks first increases (up to fields $B = 2.5 - 3T$), and then gradually decreases and is fading away in the fields $B \cong 5T$ which is similar to the behavior of the flux-flow resistance oscillations for Bi-systems (see, for example, Fig. 4 in [14]).

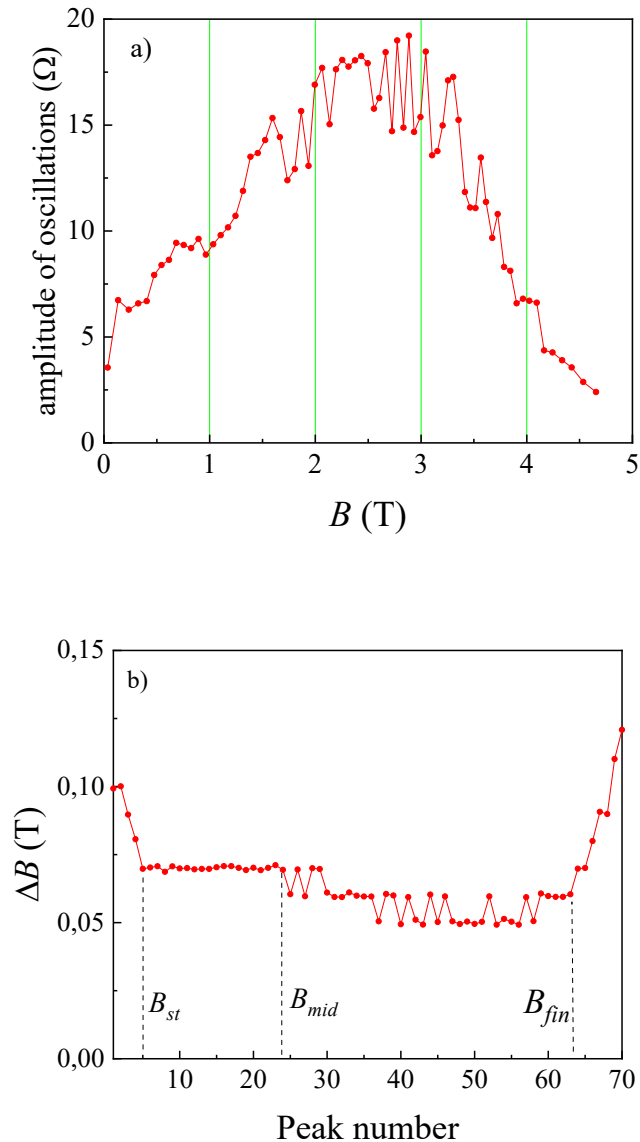

Fig. 4.5. (a) The amplitude of "oscillations" vs $B$ and (b) the width of the peaks, $\Delta B$, in dependence on the peak number

In Figure 4.5 (b), after the initial transition part (the first three peaks), at $B > B_{St}$, three regions can be distinguished on the dependence of $\Delta B$ on the magnetic field $B$ and on the peak number, $n$ (here start field, $B_{St} = 0.4$T, middle field, $B_{Mid} = 1.7$T and final one, $B_{Fin} = 3.9$T):

- $B_{St} < B < B_{Mid}$, a sufficiently large region ($n = 4 \div 24$) of periodicity with $\Delta B = 0.07T$, which, it seems, corresponds to a system of independent Josephson vortices by sizes $\lambda_{ab}$ and $\lambda_c$;

- $B_{Mid} < B < B_{Fin}$, a region with intermittent periodicity type ($n = 24 \div 64$) where sections with $\Delta B = 0.06$T, $0.05$T and then again $0.06$T and intervals between them are visible on the graph. Here the compaction of the Josephson vortex system with increasing $B$ should lead to the formation of certain types of vortex ordering (triangular lattice, square lattice, etc.). In this case, edge effects (the influence of sample boundaries) according to Koshelev [17], [18] become important;

- $B > B_{Fin}$, a region of stepped structure fading on $R_c(B)$ ($n > 64$) due to further compaction of the vortex lattice and the interaction of vortices, where multi-vortex effects become significant.

As mentioned above, in a multilayer anisotropic superconductor the structure of the individual vortices considered within the framework of a Josephson-coupled layer model based on the Lawrence-Doniach model in the London limit [8], [33], [34] (see Sec. 2.2).

In such superconductors, the magnetic penetration depths $\lambda_{ab}$ and $\lambda_c$ give the scale for the decay of the supercurrent along the plane of IJJ and across ($c$-axis), соответственно (see Fig. 2.3). They determine the "external" contour of the Josephson vortex as a region for concentration of the magnetic flux quantum, $\Phi_0$.

The value of the magnetic field required to generate one quantized Josephson vortex (the lower critical field) in such a system, according to Koshelev and Dodgson [34], is equal to (in our notation):

$$B_{c1} = \frac{\Phi_0}{4\pi\lambda_{ab}\lambda_c} \ln\left(\frac{0.44\lambda_c}{s}\right). \tag{4.2}$$

Based on the fact that each change in the field by this value leads to the birth of a new vortex, we assume to estimate $\Delta B \cong B_{c1}$, where $\Delta B$ is the observed (in the first region) period of the peak structure on $\Delta R_c(B)$ (see Fig. 4.5 b). An estimate with the parameters of Table 1 for our sample, $\lambda_{ab} = (300 \pm 15)$nm, $\gamma = 18$, gives $\Delta B \cong B_{c1} = (0.077 \pm 0.007)$T in good agreement with the experimental data, $\Delta B (= 0.07$T$)$.

Thus, we with reasonable confidence can say that the observed period, $\Delta B$, of the peak structure on the descending branch of $R_c(B)$ is associated with the sequential switching off (quenching) of Josephson vortices (one by one) as the magnetic field decreases. In fact, we have a quantum picture: flux-flow resistance "counts" Josephson vortices, each of which contains a magnetic flux quantum.

We can draw an analogy between our results and the calculations of Owen-Scalapino [43] and the corresponding experiments (see, for example, [44]) on the resonance reaction of the critical current to the emergence of each Josephson vortex as the magnetic field increases in a single long ($L \gg \lambda_J$) Josephson junction. In [43] it is shown that a periodic peak structure is formed on the $J_c(H)$ dependence, corresponding to the addition of one vortex carrying a quantum of magnetic flux. The period of this structure is $\Delta H \sim \Phi_0/\pi d \lambda_J$, where the quantity $(\pi d \lambda_J)$ corresponds to the area of the Josephson vortex in a single long Josephson junction (see section 2.1).

## 5. Conclusions

In conclusion, compare our data on the observation of the periodic stepwise structure in the flux-flow regime for NCCO with the periodic oscillations of the flux-flow resistance observed for micron sized mesostructures in Bi2212 [11], [12], [13], [14] and in other mesoscopic HTSC systems with intrinsic Josephson effect [15], [16].

In these highly anisotropic structures, the observed oscillation periods are determined by the product of the interlayer distance, $s$ (=1.5 nm), and the mesa size, $L$ (= $20 \div 30 \mu m$): $H_p = \Phi_0/2sL$ initially and doubles to $H_0 = \Phi_0/sL$ with increasing magnetic field.

This kind of behavior is typical for the resonant response of the critical current to the magnetic field in a single "small", ($L < \lambda_J$), Josephson junction (see Section 2.1). It is a such interpretation of the flux-flow oscillations in Bi2212 system that is considered by Kadowaki [14] as well as by Ustinov and Pedersen [45].

According to [14] this phenomenon for mesoscopic intrinsic Josephson junctions in single crystalline Bi2212 can be understood from the fact that the critical current in a conventional Josephson junction oscillates with a $H_0$ period in accordance with the one flux quantum entering or exiting from the junction. The oscillation phenomenon with the period of $H_0$ resembles the Fraunhofer oscillation pattern very well. Indeed, the phenomena observed have an identical oscillation period with $H_0$ in a single layer, irrespective of the multilayered stack of the intrinsic Josephson junctions.

Furthermore, Ustinov and Pedersen [45] has made a computer simulation of the Fiske steps in a single layer Josephson junction and have found an interesting behavior: the $H_0/2$ period occurs in a lower field region and the period shifts to the $H_0$ period gradually at higher fields even in a single junction. They conclude that the flux-flow voltage oscillations with two different periods in a magnetic field, observed for Bi2212 micron sized mesostructures, have their origin in the Fiske mode excitations.

Fiske steps were studied by flux-flow resistance oscillation in a *narrow* ($L \sim 1.8$ μm) stack of Bi2212 junctions by Kim et al. [46]. The current-voltage characteristics of BSCCO IJJs structure under an external magnetic field showed pronounced Fiske steps as an evidence of high frequency excitation. The alternative appearance of even and odd Fiske steps resembles the behavior of a single junction, although there are more than 100 junctions in one stack.

We are dealing with a multilayer stack of *long* JJ ($L = 800 \mu m, L \gg \lambda_J$) and the step structure we observe is caused by the system's response to a change in the number (decrease, exit one by one) of the Josephson vortices with decreasing magnetic field. Characteristic length scales, describing the magnetic size of the Josephson vortex in our anisotropic layered superconductor, correspond to the values of penetration depths $\lambda_{ab}$ and $\lambda_c$. The theoretical or technical reasons why quantized steps are observed at. the descending branch (field output) and absent at the ascending branch (field input) of $R_c(B)$ dependence now remain unclear.

**Acknowledgments**